\def\etal{{\it et~al. }}

\documentstyle{laa}
\begin{document}

\title{On the Initial Morphology of Density Perturbations}

\author{Alejandro Gonz\'{a}lez S.\inst{1,2} \&
        Oscar Mart\'{i}nez B.\inst{2}}

\institute{Astronomy Centre, MAPS, University of Sussex,
Brighton, BN1 9QH, UK
\and
Instituto Nacional de Astrof\'{i}sica, Optica y Electr\'{o}nica,
    A.P. 51 y 216, C.P. 72000, Tonantzintla, Puebla. M\'{e}xico}
\thesaurus{ }
\offprints{Alejandro Gonz\'{a}lez S.}
\date{}
\maketitle
\begin{abstract}
     The morphological distribution of primordial density peaks is assessed.
Previous determinations, those of Peacock \& Heavens (1985) and those of
Bardeen
\etal (1986) have contradictory concluded that there exist a tendency towards
prolate or towards oblate shapes, respectively. By using two methods, the
Hessian
and the inertia tensor momenta we have performed numerical determinations of
the
triaxiality of density perturbations in Gaussian random fields with power law
spectra. We show that there is no present any tendency of shapes, and that the
triaxiality distribution is independent of the spectral index. Moreover, it is
shown
that the results of Peacock and Heavens are compatible with our determinations.
These results are in complete agreement with current triaxial distributions
infered
for galaxies and clusters both, from observations and from numerical
simulations.


\end{abstract}

\section{Introduction}

   In the extensive analysis of Random Gaussian fields by Bardeen, Bond,
Kaiser \& Szalay (1986, BBKS), it was found that the density
maxima are intrinsically triaxial with some preference towards oblate shapes.
Several authors have attempted to connect such a distribution of triaxiality,
or the
height of the peaks in the primordial density field  to the observed
distribution of triaxiality of galaxies and clusters. For instance, Dubinski
(1992) based on the Peacock \& Heavens (1985, PH) results, compared the
triaxiality
distribution of density peaks to the triaxiality of dark halos in hierarchical
models. He aimed to study whether the initial triaxiality distribution, with a
tendency towards prolateness, is preserved during the non-linear evolution of
the density field. His hypothesis was, that if the shape of peaks indeed
determines
the shape of dark haloes around galaxies, an excess of prolate dark halos may
be
expected because of the natural bias for prolate initial conditions, which he
did
not find. Furthermore, Evrard (1989) and Evrard, Silk \& Szalay (1990) tried to
explain the Hubble sequence by identifying elliptical galaxies with the highest
peaks in the density field, and spirals with the lowest peaks.

   Some other authors have attempted to infer the 3-dimensional shape of
galaxies
and clusters from the observed projected axial ratios. Fasano \& Vio (1991),
for
example, analysed 204 elliptical galaxies and showed that the distribution of
triaxiality of these galaxies is incompatible with both, a distribution of pure
oblate or pure prolate spheroids. A very low number of round elliptical
galaxies
was found in their sample. Lambas, Maddox \& Loveday (1992) studied a larger
sample of 20399 galaxies, including spiral and ellipticals galaxies. They
confirmed
that elliptical galaxies are triaxial with no preference for oblate or prolate
shapes.

    Clusters of galaxies have also been studied by e.g. Carter \& Metcalfe
(1980),
who found not only that clusters of galaxies are flatter than elliptical
galaxies,
but also that their distribution of shapes is better fitted with a set of
oblate
spheroids. On the other hand, Binggeli (1982) analysed the apparent ellipticity
of
44 Abell clusters, finding that the shapes distribution is consistent with
prolate
shapes. Furthermore, Plionis, Barrow \& Frenk (1991), with a more complete
sample
of 6000 clusters determined that the morphological distribution is inconsistent
with pure oblate objects, and conclude that a triaxial distribution of shapes
would fit better their sample. Such a trend has been confirmed by using
numerical
simulations in hierarchical models by Efstathiou \etal (1985) and Frenk \etal
(1988). Concerning superclusters, there also exist contradictory claims
regarding
their shapes. West (1989) generated a catalog of 48 probable superclusters by
using
a large sample of Abell clusters and found that their shape is consistent with
prolate objects. Meanwhile, Plionis \etal (1992) by employing several
statistical
techniques and a sample of Abell and ACO (Abell, Corwin \& Olowin, 1989)
clusters,
obtained that superclusters are triaxial with a strong preference toward
oblatness.
Because the density field is in the linear regimen on large scales, it is
possible
that the morphology of large-scale perturbations shall reflect the primordial
shapes. In adition, there are theoretical evidences in favor that information
of
the initial shape at the moment of galaxy formation can be preserved after the
relaxation process (Aarseth \& Binney, 1978).

    Clearly, there is no consensus on the distribution of triaxiality for the
different observed population of cosmic objects. Whilst several statistical
properties of Gaussian and Non-Gaussian statistics have been addressed by
e.g. BBKS, Couchmann (1987), Barrow \& Coles (1987, 1990) and Coles (1990),
concerning shapes of peaks in Gaussian fields there are only two studies
with contradictory conclusions. For PH peaks present a tendency towards prolate
shapes, whereas the analytical study by BBKS gave a trend towards oblate
shapes.
The real shape and triaxial distribution of density perturbations, are highly
desirables if one aims to understand the effect of the non-linear evolution of
the
density field on the morphology of perturbations, and moreover, the effect of
tidal
interactions. In this paper, we perform numerical determinations of the
triaxiality
of density peaks and show that Gaussian random fields present no tendency for
oblate or prolate shapes. They are intrinsically triaxials. We also confirm the
BBKS statement that triaxiality distribution is not sensitive to the spectral
index
of power law spectra and therefore is useless to constraint cosmological
models. For
these goals the rest of the paper is organized as follows. In Section 2, we
start
by briefly review the theretical procedure to assess shapes. A description of
the
adopted numerical methods is done in Section 3. Section 4, contains our results
and a comparison with prior determinations. A discussion and our conclusions
are
presented in section 5.

\section{Shape of Peaks: theoretical formulation}

In the following analysis we shall adopt the procedure and notation of BBKS. To
first approximation one can expand the density field around a peak as
\begin{equation}
\delta ({\bf r})=\delta (0) + r_{i} \frac{\partial \delta ({\bf r})}
             {\partial r_{i}}\mid_{{\bf r}=0} + \frac{1}{2} r_{i}r_{j}
             \frac{\partial^{2}\delta ({\bf r})}{\partial r_{i}\partial
             r_{j}}\mid_{{\bf r}=0} + ...,
\end{equation}
where the second term on the r.h.s. is zero due to the extremum condition.
The term composed by the second derivatives of the field is the Hessian matrix
\begin{equation}
                H\equiv -\frac{\partial^{2}\delta}{\partial r_{i}\partial
r_{j}},
\end{equation}
whose eigenvalues ordered, without lack of generality, as $\lambda_{1} >
\lambda_{2} > \lambda_{3}$ represent the quadratic approximation
\begin{equation}
      \delta (r) = \delta (0) - \sum \lambda_{i} r_{i}^{2}/2,
\end{equation}
which defines an ellipsoid with semiaxes
\begin{equation}
       a_{i}= [\frac{2 \delta (0) - \delta_{t}}{\lambda_{i}}]^{1/2},
\end{equation}
where $\delta_{t}$ defines the value of the isodensity countour for which the
triaxiality is to be determined. Equation (4) is the definition adopted by PH
to
analyse the morphology of density peaks in terms of the ratio of their main
axes.

   Aditionally, the triaxiality parameters are defined by BBKS in terms of the
Hessian eigenvalues as
\begin{equation}
       \epsilon = \frac{\lambda_{1}-\lambda{3}}{2\sum \lambda_{i}},
\hspace{2.0cm}
       p = \frac{\lambda_{1} - 2\lambda_{2} + \lambda_{3}}{2\sum\lambda_{i}},
\end{equation}
where $\epsilon$ measures the ellipticity in the $\lambda_{1}-\lambda_{3}$
plane and
$p$ measures the prolatness or oblatness. If $0\geq p \geq -\epsilon$ then the
ellipsoid is prolate-like, while if $\epsilon \geq p \geq 0$ it is oblate-like.
The
limiting cases are, $p= -\epsilon$ for prolate spheroids and $p=\epsilon$ for
an
oblate spheroid.

   The conditional distribution, $P(\epsilon,p\mid x)$, which describes the
probability of finding a density perturbation with triaxiality parameters
$\epsilon$ and $p$ in the interval $\epsilon + d\epsilon$ and $p+dp$, given
the value of the normalized curvature $x=\nabla^{2}\delta/\sigma_{2}$ (see eq.
[7.6] in BBKS), is
\begin{equation}
 P(\epsilon,p\mid x)= (\frac{2}{2\pi)})^{1/2}\frac{x^{8}}{f(x)}\exp
(-(5/2)x^{2}
   (3\epsilon^{2} + p^{2}))) W(\epsilon,p),
\end{equation}
where we have used $x=x^{*}=\gamma\nu$, the most probable value of the
curvature,
and $\gamma$ is a function of the spectral index $\gamma= [(n+3)/(n+5)]^{1/2}$.
Hereafter the height of the peaks will be expressed in units of the root mean
square $\sigma$ of the density field as $\delta =\nu\sigma$. Thus, given a
probability $P$ and a threshold heigth $\nu_{max}$ of the peaks, equation (6)
defines an isoprobability curve which encloses at least $100(1-P) \%$ of all
the peaks in the density field with height $\nu < \nu_{max}$. Figure 1 exhibits
some of the isoprobability contours as a function of the probability and height
of
the peaks for the spectral indexes $n=1, -2$. Only these spectral indexes were
included to enhance any possible dependence of the triaxiality distribution on
the spectrum index. Nevertheless, no extreme dependence is observed. Our
numerical
results will be compared with the theoretical contours of Figure 1.

\begin{figure*}      
\vspace{17.5cm} 
\caption{Isoprobability contours of the distribution of triaxiality for peaks
         as a function of their height $\nu$, for $n=1, -2$. The theoretical
$50\%$
         isoprobability contours are included, which shall contain half of the
         total number of peaks.}
\end{figure*}

\section{Numerical Approximation}

    We have used the Hessian and the inertia tensor methods to assess the
intrinsic
shape of peaks. THe first one, was used by PH to carry out numerical
determinations
of the morphology of peaks, whereas the second method was adopted by BBKS for
their
analytical treatment. Since we aim to numerically test and collate these
results,
both methods require of smoothing the discrete density field, randomly
generated in
a $64^{3}$ box, and an interpolation process. We based this latter on the
triangular-shaped cloud method (Hockney \& Eastwood, 1988). This enables us the
reconstruction of the density field at any point within the cubic grid.
Further, in
order to apply the interpolation, it is also required to set up the minimum
number
of contributing points-density to the inertia momenta necessary to get stable
results of the morphology. In the calculations, we have associated a unit mass
to
each point of the peak, though various others weighting schemes can be used,
see
e.g.  West (1989b) and Plionis \etal (1992). A fully detailed description of
the
method used, and the determination of the minimum number of grid points and
steps
of interpolation can be found in Gonz\'{a}lez (1994). $N_{s}$ denotes the
number of
sub-divisions of each $1^{3}$ cube that is crossed by the isodensity surface.

\subsection{Methodology}

We generated random fields with a power law spectrum, $P(k)=Ak^{n}$, with
spectral
indexes $n= 1,0,-1,-2$. The $64^{3}$ box used in the calculation has a physical
length of $64h^{-1}$Mpc on a side. We then proceed as follows:
\begin{enumerate}
\item The density field is convolved with a Gaussian filter
\begin{equation}
      W_{G}=\exp (-(k^{2} R^{2})/2),
\end{equation}
      where $k$ is the perturbation wavenumber and $R$ is the filtering radius.

\item The positions of grid maxima are found, which are defined as those grid
      points with overdensity $\delta_{max}$ higher than the density of their
      26 nearest neighbours. In general, the grid maxima positions do not
      coincide with position of real peaks.

\item The value of the isodensity surface $\delta_{t}$, for which the
triaxiality
      parameters will be calculated, is chosen by introducing a factor $f$
\begin{equation}
      \delta_{t} = f \delta_{max},
\end{equation}
      with $0 << f \leq 1$. Then, the inertia tensor is calculated
\begin{equation}
      I_{kl}=\frac{1}{N}\sum_{i=1}^{m} [(x_{k}-<x_{k}>)(x_{l}-<x_{l}>)],
\end{equation}
which involves the use of the grid points and the interpolation code of
$N_{s}=14$
steps, and where $k,l=1,2,3, x_{1i}, x_{2i}$ and $x_{3i}$ are the Cartesian
coordiantes of the $ith$ point. For a given $f$, the sum runs over all the
cubic
nodes with $\delta \leq \delta_{t}$. The inertia tensor eigenvalues $a_{3}^{2}
>
a_{2}^{2} > a_{1}^{2}$, are then related to the eigenvalues of the Hessian
matrix
\begin{equation}
\lambda_{1}=1/a_{1}^{2},\hspace{.5cm} \lambda_{2}=1/a_{2}^{2} \hspace{.5cm}
            {\rm and}\hspace{.5cm} \lambda_{3}=1/a_{3}^{2}.
\end{equation}
\item In order to assure the validity of the quadratic approximation, the
calculation of the Hessian matrix must be made close to the position of the
real peak. Therefore, one of our requeriments of the inertia tensor method, is
to perform an interpolation process to localize the real maximum. Afterwards,
the
Hessian matrix is constructed by interpolating the density at points separated
at distances $d=R/3, R/2, R$ from the peak center. These three distances were
used to test the numerical stability and constancy of the shape around peaks.
\end{enumerate}

\section{Results}

\subsection{Statistical analysis}

  Figure 2 shows the distribution of the triaxiality parameters as a function
of the threshold height. It confirms the trend for the highest peaks, $\nu >
3.5$,
to be more spherical than the lower ones as found by BBKS. The large dispersion
of
peaks shapes for all over the $\epsilon - p$ plane, comes from the peaks lower
than
$\nu < 1.5\sigma$. Meanwhile,  the distribution of shapes for peaks in the
interval $1.5< \nu < 3$, --where the number density of peaks is maximum--, is
more
concentrated towards sphericity and at least $50\%$ of them lay within the
theoretical isoprobability contours. If we look at the form of these
isoprobability
contours, we will notice that the major part of the area bounded by them, is
into
the oblate shapes side. In order to check whether this should be, or not,
understood as a tendency towards oblateness, we have calculated the ratio
$\Gamma$
of the number of oblate ellipsoids $N_{o}$ to the number of prolate ones
$N_{p}$,
both of which are numerically obtained. Any significant trend must be reflected
in
$\Gamma$. In the second and third columns of Table 1 we display this ratio for
different spectral indexes as a function of the height of peaks. It is easily
oberserved  that there is not any morphological preference. Moreover, when
we only consider the highest peaks no strong trend is detected, even when they
constitute a small sample.
\begin{figure*}      
\vspace{8.5cm}         
\caption{Morphology of peaks as a function of their heigh, for $n=1$. The
         theoretical $50\%$ isoprobability contours is included, which
         contains approximately half of the total number of peaks. The mean
         triaxiality parameters are $<\epsilon>=0.174$ and $<p>=2.3\times
         10^{-2}$.}
\end{figure*}

   The histogram of Figure 3 displays the distribution of oblate and prolate
ellipsoids. This histogram and the ratios of Table 1, suggest that if an
especific
trend of shapes exists, in fact it behaves stochastically, and has no
statistically relevance. Therefore, an intrinsic triaxial shape of density
perturbations would be the most acceptable conclusion. An inspection of the
distributions of Figure 3 also confirms that the results
are not significantly dependent on the spectral index. The distribution of
ellipticity and prolateness for all the models exhibit a slight tendency for
oblate shapes. The mean values $<\epsilon > = 0.18\pm 0.02$ and $<p>= (.13
\pm 0.003)\times 10^{-3}$ incorporate all the models. These values and the
distribution of peaks in the $\epsilon -p$ plane show disagreement both with
the
theoretical predictions of BBKS and with the numerical ones of PH. In order to
compare our results with those of these latter authors, it is necesary to
assess
the shapes by resorting to the Hessian matrix as indicated in section 3. Once
the
Hessian is diagonalized we take the ratio of axes defined by equation (10)
\begin{equation}
    \left(\frac{\lambda_{3}}{\lambda_{1}}\right)^{1/2}=
    \frac{a_{1}}{a_{3}}=\frac{{\rm short-axes}}{{\rm long-axes}}=s,
\end{equation}
\begin{equation}
    \left(\frac{\lambda_{3}}{\lambda_{2}}\right)^{1/2}=
    \frac{a_{2}}{a_{3}}=\frac{{\rm middle-axes}}{{\rm long-axes}}=m.
\end{equation}
\begin{table*}                     
\begin{center}
\begin{tabular}{|c||p{2.5cm}|c|p{2.5cm}|}
\hline
\multicolumn{4}{|c|}{Shapes of Density peaks} \\
\hline\hline
\multicolumn{1}{|c||}{Spectral index }     &
\multicolumn{1}{|c|}{$1^{st}$Realization}  &
\multicolumn{1}{|c|}{$2^{nd}$Realization}  &
\multicolumn{1}{|c|}{$3^{th}$Realization}  \\
\multicolumn{1}{|c||}{n}                   &
\multicolumn{1}{|c|}{(Inertia Tensor)} &
\multicolumn{1}{|c|}{(Inertia Tensor)} &
\multicolumn{1}{|c|}{(Hessian)} \\
\multicolumn{1}{|c||}{}                   &
\multicolumn{1}{|c|}{$\Gamma=N_{o}/N_{p}$} &
\multicolumn{1}{|c|}{$\Gamma=N_{o}/N_{p}$} &
\multicolumn{1}{|c|}{$\Gamma=N_{o}/N_{p}$} \\
\hline
$n= 1$        &                &              &             \\
$\nu > 3.5$   & $110/105$      & $112/111$    & $104/101$   \\
$\nu > 2.5$   & $499/495$      & $502/506$    & $503/498$   \\
$\nu > 1  $   & $3140/3144$    & $3093/3092$  & $3098/3101$ \\
$\nu > 0  $   & $3283/3276$    & $3337/3341$  & $3305/3306$ \\
\hline
$n= 0$        &                &              &             \\
$\nu > 3.5$   & $93/95$        & $113/109$    & $107/102$   \\
$\nu > 2.5$   & $413/409$      & $402/410$    & $423/417$   \\
$\nu > 1  $   & $2527/2512$    & $2314/2321$  & $2323/2321$ \\
$\nu > 0  $   & $2687/2693$    & $2498/2475$  & $2548/2539$ \\
\hline
$n=-1$        &                &              &             \\
$\nu > 3.5$   & $70/68$        & $73/76$      & $69/72$     \\
$\nu > 2.5$   & $297/305$      & $305/308$    & $313/309$   \\
$\nu > 1  $   & $1789/1797$    & $1810/1797$  & $1805/1812$ \\
$\nu > 0  $   & $2057/2048$    & $2023/2026$  & $2063/257$  \\
\hline
$n=-2$        &                &              &             \\
$\nu > 3.5$   & $37/32$        & $42/41$      & $38/41$     \\
$\nu > 2.5$   & $153/158$      & $165/163$    & $157/155$   \\
$\nu > 1  $   & $1071/1083$    & $1112/1096$  & $1086/1083$ \\
$\nu > 0  $   & $1386/1394$    & $1367/1356$  & $1352/1343$ \\
\hline
\end{tabular}
\end{center}
\caption{Ratio of the number of oblate objects to the number of prolate one, as
         a function of the spectral index $n$ and the high of peaks. The used
        method is indicated.}
\end{table*}
Figure 4 is an example of the distribution of the ratio of axis in the $s-m$
plane. A visual judgement of such a distribution can easily lead us to wrongly
conclude --as PH did-- that a tendency of the peaks exists towards prolateness.
In a $s-m$ diagram,  a difference between a slight tendency to oblatness and a
slight tendency to prolateness would be quite clear and not neglegible. The
reason for this artificial discrepancy with our results is that PH did not
consider the line which separates these two tendencies: $p=0,\hspace{.2cm}
\forall\hspace{.12cm} \epsilon$ in the interval $0\leq\epsilon\leq 0.33$, i.e.
all the eigenvalues which satisfy
\begin{equation}
\lambda_{1}- 2\lambda_{2} + \lambda_{3}= 0,
\end{equation}
or in terms of the ratio of axes
\begin{equation}
   \frac{1}{s^{2}} - \frac{2}{m^{2}} + 1 = 0,
\end{equation}
which gives the equation of the curve
\begin{equation}
        s=\frac{m}{\sqrt{2-m^{2}}}.
\end{equation}
When this function is included in the $s-m$ diagram of Figure (4), it is
observed
that the results of PH are also consistent with no tendency, perturbations are
intrinsically triaxial as is further shown both by the distribution of the
number
of oblate and prolate shapes of Figure 5 and by the fourth column of Table 1.
Going
back to the results of BBKS one would wonder why there should be a {\it natural
bias} for oblate shapes ?

\begin{figure}             
\centering   
\vspace{6.5cm}
\caption[]{Distribution of oblate and prolate shapes of peaks as a function of
the
           spectral indexes, $n=1, -2$, and hight $\nu > 0$. These
distributions
           correspond with the cases of Table 1. The distribution in dashed
line is
           a theoretical distribution of triaxiality, constructed by generation
of
           random triplets $\lambda_{i}$ (see Sec. 4).}
\end{figure}

\subsection{Random generation of $\lambda_{i}$}

The primordial density field is a Gaussian random field, i.e. it is
generated by superposing a large number of Fourier modes with phase angles
drawn at random from a uniform distribution in the interval $0-2\pi$. The
shapes
of density peaks, regarding them since the Hessian point of view, are the
result
of the curvatures of the density field on the three main axes. The eigenvalues
of the Hessian matrix are therefore randomly generated numbers as well. The
value of one does not affect the value of the other two. Under this assumption,
it is no clear how a tendency towards oblateness can arise. It is also under
this assumption that we have generated triplets of random numbers, without
paying attention for them to fulfill the basic requirements of a cosmological
density field; power law spectra, spectral index, etc. This has the goal of
trying to reproduce our triaxiality distribution as a random process. This
would indicate that the tendency infered from the BBKS results is a consequence
of several analytical simplifications which hide the real triaxial shape of
peaks.

\begin{figure}             
\centering   
\vspace{8cm}
\caption[]{Distribution of peaks according with the ratio of axis, $s$
           and $m$, as presented by Peacock and Heavens (1985), where an
           artificial tendency towards prolate shapes is observed.}
\end{figure}

\begin{figure*}            
\vspace{8.5cm}   
\caption{a) The same as Figure 4 for $n=-1$, but here the boundary curve (eq.
         15) between the oblate and the prolate shapes is included. b) This
         distribution of oblate and prolate shapes, and fourth column of Table
         1, show that the intrinsic shape of peaks is triaxial}
\end{figure*}
We will call the triplets, eigenvalues,such that $0 < \lambda_{i} \leq 1$.
Once they are put in an increasing order, we determine the triaxiality
parameters
through equation (5). The results of this are also displayed in the
distribution of
Figure 3 (in dashed line), which present no substantial differences with the
distribution of shapes for a Gaussian random field obtained by using the
inertia
tensor and Hessian methods. Moreover, we can provide a meassure of how equal
these
two distributions are, in terms of their moments; mean $<p>$, standard
deviation
$\sigma$, skewness $s$ and kurtosis $k$. These moments characterize the
observed
asymmetry of the triaxiality distribution. For the Gaussian density field we
obtain:
$\bar{p}=1.324\times 10^{-2}, \sigma=0.1189, s= 0.9091$ and $k=0.9364$. Whereas
for the distribution determined by the random triplets we get: $<p>=8.53\times
10^{-3}, \sigma=0.1194, s=0.9251$ and $k=0.9635$. Therefore, it is quite likely
that
this two distributions are generated by the same stochastic process.

\section{Conclusions}

The present investigation underlines the nature of primordial density
peaks and the dependence of the triaxiality distribution on the spectral
index, for power law spectra. This numerical analisys explored the
veracity of the PH and BBKS conclusions on these points.

For all the spectral indexes, we confirmed the BBKS result that high peaks
tend to be more spherical than the lower ones. When the whole sample of
peaks is considered, an intrinsic triaxial morphology is observed. This result
solves a controversy concerning a natural initial bias towards prolateness
shapes claimed by Heavens and Peacock (1985) and Dubinski (1992), or towards
oblateness infered from the Bardeen \etal (1986) results. Moreover, the recent
determinations on the intrinsic shape of elliptical galaxies by Fasano \& Vio
(1991)
and Lambas \etal (1992), which resulted consistent with triaxial objects,
suggest
that the distribution of shapes survives the non-linear evolution.

We shall study the aspherical collapse of perturbations in the non-linear
density field, investigating the relationship between the initial asphericity
and the stability of properties of the distribution of collapse structures in
the presence of a tidal field and in isolation

\section {Acknowledgements}

This paper presents part of the AGS Ph.D. thesis at the University of Sussex.
AGS acknowledge Peter Thomas for his supervision of this work, and thanks
him and L. Moscardini for useful comments. Almost all the calculations were
carried out at the Starlink Sussex node. OMB thanks a Ph.D. grant No. 86613
from CONACYT at M\'{e}xico.

\section{References}
\par\noindent
{Aarseth, S.J., Binney J.,1978, MNRAS 185, 227}
\par\noindent
{Bardeen J.M.,, Kaiser N., Bond R., Szalay R, 1986, ApJ 304, 15}
\par\noindent
{Barrow J. D., Coles P., 1987, MNRAS 228, 407}
\par\noindent
{Barrow J. D., Coles P., 1990, MNRAS 244, 557}
\par\noindent
{Binggeli B., 1982, A\&A 107, 338}
\par\noindent
{Carter D., Metcalfe N., 1980, MNRAS 191, 325}
\par\noindent
{Coles P., 1989, MNRAS 238, 319}
\par\noindent
{Couchmann H. M. P., 1987, MNRAS 225, 777}
\par\noindent
{Evrard A. E., 1989, ApJ 341, 26}
\par\noindent
{Evrard A.E., Silk J., Szalay A., 1990, ApJ 365, 13}
\par\noindent
{Dubinski J., 1992, ApJ 401, 441}
\par\noindent
{Fasano G. \& Vio R., 1991, MNRAS 249, 629}
\par\noindent
{Efstathiou G., White S.D.M., Frenk C.S., Davis M., 1985, Nature 317, 595}
\par\noindent
{Frenk C.S., White S.D.M., Davis M., Efstathiou G., 1988, ApJ 227, 507}
\par\noindent
{Gonz\'{a}lez S. A., 1994, Ph. D. Thesis, Sussex University}
\par\noindent
{Lambas D.G., Maddox S., Loveday J., 1992, MNRAS 258, 404}
\par\noindent
{Peacock J.A., Heavens A. F., 1985, MNRAS 217, 805}
\par\noindent
{Plionis M., Barrow J. D., Frenk C. S., 1989, MNRAS 249, 662}
\par\noindent
{Plionis M., Valdarnani R., Jing Y., 1992, ApJ 398, 12}
\par\noindent
{West M., 1989, ApJ 347, 610}
\par\noindent
{West, M., 1989b, ApJ 344, 535}
\end{document}